\documentstyle[eqsecnum,aps,prd]{revtex}
\begin{document}
\newcommand{\gsim}{\mbox{\raisebox{-1.0ex}{$\stackrel{\textstyle >}
{\textstyle \sim}$ }}}
\newcommand{\lsim}{\mbox{\raisebox{-1.0ex}{$\stackrel{\textstyle <}
{\textstyle \sim}$ }}}
\newcommand{\qc}{\tau_{\rm QC}}
\newcommand{\qd}{\tau_{\rm QD}}
\newcommand{\lt}{\tau_{\rm life}}
\newcommand{\bfx}{{\bf x}}
\newcommand{\bfy}{{\bf y}}
\newcommand{\bfr}{{\bf r}}
\newcommand{\bfk}{{\bf k}}
\newcommand{\bkp}{{\bf k'}}
\newcommand{\order}{{\cal O}}
\newcommand{\beq}{\begin{equation}}
\newcommand{\eeq}{\end{equation}}
\newcommand{\beqa}{\begin{eqnarray}}
\newcommand{\eeqa}{\end{eqnarray}}
\newcommand{\lmk}{\left(}
\newcommand{\rmk}{\right)}
\newcommand{\lkk}{\left[}
\newcommand{\rkk}{\right]}
\newcommand{\lbk}{\left\{}
\newcommand{\rbk}{\right\}}
\newcommand{\call}{{\cal L}}
\newcommand{\calh}{{\cal H}}
\newcommand{\phic}{\phi_c}
\newcommand{\pphi}{p_{\phi_c}}
\newcommand{\ppp}{\partial}
\newcommand{\rdm}{\rho_r}
\newcommand{\deco}{\delta_{QD}}
\thispagestyle{empty}
\thispagestyle{empty}
{\baselineskip0pt
\leftline{\large\baselineskip16pt\sl\vbox to0pt{\hbox{Department of Physics}
               \hbox{Kyoto University}\vss}}
\rightline{\large\baselineskip16pt\rm\vbox to20pt{\hbox{KUNS-1360}
               \hbox{\today}
\vss}}%
}
\vskip15mm
\begin{center}
{\large\bf Quantum Decoherence of Subcritical Bubble in Electroweak
Phase Transition}
\end{center}
\begin{center}
{\large Tetsuya Shiromizu} \\
\sl{Department of Physics, Kyoto University, Kyoto 606-01, Japan}
\end{center}
\begin{center}
{\it to appear in Progress of Theoretical Physics}
\end{center}
\begin{abstract}
In a weakly first order phase transition the typical scale
of a subcritical bubble calculated in our previous papers turned
out to be too small. At this scale quantum fluctuations may dominate
and our previous classical result may be altered. So we examine the critical
size
of a subcritical bubble where quantum-to-classical transition occurs through
quantum decohere
nce. We show that
this critical size is almost equal to the typical scale which we
previously obtained.
\end{abstract}
\vskip1cm

\baselineskip25pt

\section{Introduction}

Non-equilibrium electroweak phase transition is crucial for
successful electroweak baryogenesis\cite{ctm}. So far, there
are three aspects for clarifying the structure
of this first order phase transition --1)calculations of
higher loop corrections\cite{loop}, 2)lattice calculations\cite{lattice}
and 3)subcritical bubbles\cite{gl} --.
The first two aspects  are necessary for
the quantitative construction of the potential. However,
the last aspect is the most important for clarifying the
non-equilibrium nature of the phase transition.

In this paper therefore we shall
discuss the phase transition with a given potential and
concentrate on the third aspect. Existence and nature of
supercooling is clarified through
the strength of the thermal fluctuation in the symmetric
phase. In a familiar example, thermal fluctuations yield
``bubbles" in boiled water. The bubbles perpetually repeat
expansion and collapse by strong surface tensions.
If the occupation ratio
is too large, further critical bubble cannot be created even if
the potential has barrier between two vacua. A model
of the thermal fluctuation has been first proposed by Gleiser et al\cite{gl}.
They assumed $O(3)$-symmetric configuration with the spatial scale of
the order of the correlation length. After this work
fundamental problems have been actively
investigated\cite{another}\cite{smy1}\cite{smy2}.

Recently, we have estimated the typical
size and the strength of the thermal fluctuation in the minimal
standard model by using the
subcritical bubble of $O(3)$-symmetric configuration and
statistical averaging method\cite{smy1}\cite{smy2}.
The typical size of the bubble turns out to be small compared with the
correlation
length and thus the strength of the thermal fluctuation
becomes large. The conclusion is
that the electroweak phase transition is first order one {\it without
supercooling} and therefore the ordinary
electroweak baryogenesis cannot work.

However, we must worry about the smallness of the bubble because,
as we will see soon, the number of states inside a bubble
calculated in the thermal state is $\sim {\cal O}(1)$ at critical
temperature. This might imply that the classical treatment is incomplete.
So we must estimate the critical size
where quantum-to-classical transition occurs.
(If the size is small the bubble is quantum and if large it is
classical).

The rest of the present paper is organized as follows. In Sec. II,
we review our previous study in which
the typical scale of a bubble is estimated and we point out
that the number of states inside a bubble is too small for its
classicality. Further, we estimate the life time of
subcritical bubbles. In Sec. III we derive the master equation for the
reduced density matrix to discuss  the classicality of a bubble.
In Sec. IV,
we give the lower bound of the radius for classicality
comparing these time scales and show that the critical size is the same order
as the previous one. Finally, we give
a summary and discussion in Sec. V.

Hereafter the concrete values will be calculated assuming the
Higgs mass
is 60GeV and the temperature is the critical one at which two vacua
degenerate.

\section{The Typical Size and the Life Time of Subcritical Bubbles}

We review our estimation of the typical size of the thermal
fluctuation and estimate the mean life time of the subcritical bubbles.
The Lagrangian of the Higgs field is given by
%
\begin{equation}
L =\int d^3x \lkk \frac{1}{2} \ppp_\mu \phi \ppp^\mu \phi -D(T^2-T_2^2)\phi^2
+ET \phi^3-\frac{1}{4}\lambda_T \phi^4 \rkk.
\end{equation}
%

Around the critical temperature($T_c$) the ansatz,
%
\begin{equation}
\phi(x)=\phi_+{\rm exp}\lkk -\frac{r^2}{R^2(t)} \rkk,
\end{equation}
%
is reasonable because $\phi_+$ is the asymmetric value of the field
which is most expectable value. Inserting this into the
original Lagrangian,
%
\begin{equation}
L_{\rm eff}(R, {\dot R})= L[\phi = \phi_+e^{ -\frac{r^2}{R^2(t)}}],
\end{equation}
%
we obtain the Hamiltonian,
%
\begin{eqnarray}
H_{\rm eff}(R,P) & := & P{\dot R}-L_{\rm eff} \nonumber \\
  & = &  \frac{1}{2M} P^2+\frac{2}{5}M
+\frac{1}{3}M{\alpha}(T)R^2 \nonumber \\
& =:& \frac{1}{2M}P^2+V(R),
\end{eqnarray}
%
where $P$ is the canonically conjugate momentum of $R$ and
$ M (T,R):=\frac{15{\pi}^{3/2}{\phi}_+^2R}{8{\sqrt {2}}} $.
In the case of the minimal standard model, $\alpha$ becomes
%
\begin{equation}
\alpha_{\rm ew} (T) =
\frac{4}{5}D(T^2-T_2^2)
-\frac{8{\sqrt {2}}}{15{\sqrt {3}}}ET{\phi}_++\frac{1}{10{\sqrt {2}}}
{\phi}_+^2{\lambda}_T,
\end{equation}
%
where $D,~E$ and $\lambda_T$ are determined by one-loop corrections
of electroweak particles, respectively;$D \sim 0.17$, $E \sim 0.01$ and
$\lambda_{T_c} \sim 0.035$. Further, $\phi_+=
\frac{3ET}{2{\lambda}_T}
[1+{\sqrt {1-\frac{8{\lambda}_TD}{9E^2T^2}(T^2-T_2^2)}} ] \sim 51.4$
GeV and
$T_2 \sim 92.7$GeV.

In the high temperature phase($T \geq T_c$) the thermal average of $R$ is given
by
%
\begin{eqnarray}
\langle R \rangle_T & := & \frac{{\int}dPdRR{\exp
{[-\frac{H_{\rm eff}}{T}]}}}{{\int}dPdR
{\exp {[-\frac{H_{\rm eff}}{T}]}}} \nonumber \\
                & = & \frac{{\int}^{\infty}_0dRR^{3/2}{\exp
                   {[-\frac{2M}{5 T}-\frac{M{\alpha}_{\rm ew}}{3T}R^2 ]}}}{
                    {\int}^{\infty}_0dRR^{1/2}
                    {\exp {[ -\frac{2M}{5T}-\frac{M
                    {\alpha}_{\rm ew}}{3T}R^2 ]}}} \nonumber \\
& \simeq & \frac{2{\sqrt {2}}}{{\pi}^{3/2}}\frac{T}{{\phi}_+^2}
\leq \ell (T):={\sqrt {2D(T^2-T_2^2)}},
\end{eqnarray}
%
where $\ell (T)$ is the correlation length. One can easily see that
the averaged radius
is smaller than the correlation length.

As the Gaussian ansatz is imposed on the Higgs field,
the number of the state inside a bubble of this radius becomes
%
\begin{equation}
n(T) \sim  ({\pi}^{1/2}  \langle R \rangle_T)^3
{\int}^{\infty}_0
\frac{d^3{\bf k}}{(2 \pi)^3}
\frac{1-{\exp{[-\frac{\bfk^2 \langle R \rangle_T^2}{4}]}}}{{\exp{[\frac{{\sqrt
{\bfk^2+m^2}}}
{T}]}}-1}
\end{equation}
%
in the thermal state.
Around $T=T_c$ the number is $\sim {\cal O}(1)$ and it might imply
that our classical treatment is not complete.

Next, let us estimate the life time of the subcritical bubble.
This time scale will be compared with the decoherence time in Sec. IV.
As
%
\begin{eqnarray}
MV^2 & = & PV=\frac{d}{dt}(PR)-\frac{dP}{dt}R \nonumber \\
     & = & \frac{d}{dt}(PR)+ \lmk V'(R)-\frac{1}{R} \frac{1}{2}MV^2
           \rmk R
\end{eqnarray}
%
holds from the Euler-Lagrange equation,
we obtain the virial relation by taking the long time average on
this equation;
%
\begin{equation}
{\overline {\frac{1}{2}MV^2 }}={\overline {\frac{2}{15}M+
\frac{1}{3} \alpha_{\rm ew}(T)MR^2 }}.
\end{equation}
%
Thus the mean life time of subcritical bubbles with radius $R$ is
given by
%
\begin{equation}
\tau_{\rm life} \simeq R \lkk  \frac{4}{15}
+\frac{2}{3}\alpha_{\rm ew}R^2 \rkk^{-1/2}.
\end{equation}
%
As $(2/3)\alpha_{\rm ew}R^2 \ll 4/15   $ at $T = T_c$,
$ \tau_{\rm life} \sim ({\sqrt {15}}/2)R$ holds approximately.

\section{The Derivation of Master Equation}

In the previous section, we find that the number of states is too small for
classicality and therefore classical treatment may not be complete.
However, this aspect of the number of states is not complete to determine
whether the system is quantum or classical.
In general, there are two classicality conditions:
classical correlation and quantum decoherence. The former condition
is satisfied in the case when the sharp orbit in the phase space
exists -- for example when WKB approximation is good. Unfortunately,
one cannot take the limit $\hbar \to 0$ now, otherwise
one cannot discuss the temperature dependent phase transition and
the thermal fluctuation vanishes.
One should remember the fact that the first order type effective
potential was obtained by calculating loop corrections.
Therefore, we study much elaborate determination
based on the quantum decoherence\cite{deco}.

In this section we derive the master equation for the reduced
density matrix.
For simplicity, we consider the following action of a singlet
Higgs field $\phi$;
%
\begin{eqnarray}
S[\phi, \psi, {\overline \psi}]& = & S[\phi]+S_f[\psi, {\overline \psi}]
+S_{\rm int}[\phi, \psi, {\overline \psi} ] \nonumber \\
& = & \int d^4x \lkk \frac{1}{2}({\partial}_{\mu} \phi)^2
-\frac{1}{2}m^2
{\phi}^2 -\frac{1}{4!}\lambda {\phi}^4+ i {\overline {\psi}}
{\gamma}^{\mu} {\partial}_{\mu} \psi
-f \phi  {\overline {\psi}} \psi \rkk,
\end{eqnarray}
%
where $\psi$ is the top quark which plays the role of environment.
The reduced density matrix can be written by
%
\begin{equation}
\rho_r[\phi_f,{\phi'}_f;t]=\int {\cal D}\phi_i \int {\cal D} \phi'_i
J[\phi_f,{\phi'}_f;t|\phi_i,{\phi'}_i;0] \rho_r[\phi_i,{\phi'}_i;0],
\end{equation}
%
where
%
\begin{equation}
J[\phi_f,{\phi'}_f;t|\phi_i,{\phi'}_i;0]=\int^{\phi_f}_{\phi_i}
{\cal D} \phi
\int^{\phi'_f}_{\phi'_i}{\cal D}\phi' {\rm exp}\lkk i \lbk
S \lkk\phi \rkk-S \lkk \phi' \rkk \rbk \rkk F \lkk \phi, \phi' \rkk
\end{equation}
%
and
%
\begin{eqnarray}
F[\phi, \phi' ] & = & \int {\cal D}\psi_f \int {\cal D} {\overline \psi}_f
 \int {\cal D}\psi_i \int {\cal D} {\overline \psi}_i
 \int {\cal D}\psi'_f
\int {\cal D}{\overline \psi'}_f
 \int {\cal D}\psi'_i
\int {\cal D} {\overline \psi'}_i
\rho_0[\psi_i ,{\overline \psi}_i, \psi'_i , {\overline \psi'}_i;0 ]
\nonumber \\
& \times & \int^{\psi_f}_{\psi_i} {\cal D}\psi
\int^{{\overline \psi}_f}_{{\overline \psi}_i}{\cal D} {\overline \psi}
\int^{\psi'_f}_{\psi'_i} {\cal D}\psi'
\int^{{\overline \psi'}_f}_{{\overline \psi'}_i}{\cal D} {\overline \psi'}
{\rm exp} \lkk i \lbk S_f \lkk \psi ,{\overline \psi} \rkk
+S_{\rm int} \lkk \phi, \psi, {\overline \psi} \rkk
-S_f [ \psi', {\overline \psi'} ]
-S_{\rm int} [ \phi', \psi', {\overline \psi'} ]  \rbk \rkk \nonumber \\
& = & {\rm exp} \lkk i \lbk \delta S \lkk \phi \rkk
-\delta S[\phi' ]-2\int d^4x \int d^4x' A(x-x') \phi_c(x')
\phi_\Delta(x)
+\frac{i}{2}\int d^4x \int d^4x' B(x-x')
\phi_{\Delta}(x)\phi_{\Delta}(x') \rbk \rkk.
\end{eqnarray}
%
We used the following notation:
$X_{\Delta}=X - X'$ and $X_C=(X+ X')/2$. Further, we assumed that
the initial density matrix can be written by
%
\begin{eqnarray}
\rho[\phi_i, \psi_i, {\overline \psi}_i,
\phi'_i, \psi'_i, {\overline \psi'}_i ;0]& = & \rho_0[\psi_i,
{\overline \psi}_i, \psi'_i, {\overline \psi'}_i ;0 ]
\rho_r[\phi_i, \phi'_i;0] \nonumber \\
& = & e^{-\beta H_f}  \rho_r[\phi_i, \phi'_i;0],
\end{eqnarray}
%
where $H_f$ is the Hamiltonian of the top quark.
The above $F[\phi, \phi' ]$ is called as influence functional
\cite{feynman}. Here
kernels $A(x-x')$ and $B(x-x')$ are calculated from one-loop
diagrams, respectively;
%
\begin{equation}
A(x-x')=f^2{\rm Im}[S(x-x')S(x'-x)]\theta(t-t')
\end{equation}
%
and
%
\begin{equation}
B(x-x')=f^2{\rm Re}[S(x-x')S(x'-x)],
\end{equation}
%
where $S(x-x')$ is the dressed Green's function of top quark which
has the expression
%
\begin{eqnarray}
S({\bf p},t) & = & \frac{e^{-{\Gamma}|t|}}{2{\omega}_p}
\Bigl[ \Bigl(-{\gamma}^0{\omega}_p{\rm sign}(t)+\vec{\gamma} \cdot
\vec{p} \Bigr) f(-{\omega}_p+i{\Gamma}) e^{-i{\omega}_p|t|}  \nonumber \\
& - & \Bigl({\gamma}^0{\omega}_p{\rm sign}(t)+\vec{\gamma} \cdot
\vec{p} \Bigr) f({\omega}_p+i{\Gamma}) e^{i{\omega}_p|t|} \Bigr]
\end{eqnarray}
%
and ${\Gamma}$ is the decay width given by
${\Gamma} \simeq \frac{f^2}{8 \pi} T$ in the high temperature limit.
$\delta S[\phi]$ gives the loop correction to the original potential
and $f(x)$ is Fermi distribution($f(x)=1/(e^{\beta x}+1)$).
The above calculation is almost the same as that of one dimensional
system with harmonic oscillators\cite{cd}. Also, the calculation in the
influence functional is the almost same as the one in the in-in
formalism\cite{smy2}.

Adopting the ansatz
%
\begin{equation}
\phi(x)=\phi_+{\rm exp}\lkk -\frac{r^2}{R^2(t)} \rkk
\end{equation}
%
and
%
\begin{equation}
\phi'(x)=\phi_+{\rm exp}\lkk -\frac{r^2}{{R'}^2(t)} \rkk,
\end{equation}
%
the propagator of the reduced density matrix becomes
%
\begin{eqnarray}
J(R_f,R'_f;t|R_i, R'_i;0) & = & {\rm exp} \Bigl[ i \Bigl\lbrace
\int^t_0 ds [  \frac{1}{2}M(R) \lmk \frac{dR}{ds} \rmk^2 +V(R) ]
-\int^t_0 ds [  \frac{1}{2}M(R') \lmk \frac{dR'}{ds} \rmk^2 +V(R') ]
\nonumber \\
& - & 2 \int^t_0 ds \int^t_0 ds' \int \frac{d^3 \bfk}{(2 \pi)^3}
X_\bfk (s) X_\bfk (s') [ {\cal A}(\bfk, s-s')\theta (s-s')
-\frac{i}{4}{\cal B}(\bfk, s-s') ] \Bigr\rbrace \Bigr],
\end{eqnarray}
%
where
%
\begin{equation}
{\cal A}(\bfk, s-s'):=f^2\frac{{\pi}^3}{2}{\phi}_+^2
{\int} \frac{d^3{\bf p}}{(2 \pi)^3}{\rm Re}\Bigl[ iS_{\alpha \beta}
({\bf p}, s-s')S^{\beta \alpha}({\bf p}-{\bf k}, s'-s) \Bigr],
\end{equation}
%
%
\begin{equation}
{\cal B}({\bf k}, s-s')  :=  f^2\frac{{\pi}^3}{2}{\phi}_+^2
{\int} \frac{d^3{\bf p}}{(2 \pi)^3}{\rm Im}\Bigl[ iS_{\alpha \beta}
({\bf p}, s-s')S^{\beta \alpha}({\bf p}-{\bf k}, s'-s) \Bigr]
\end{equation}
%
and
%
\begin{equation}
X_\bfk (s):=R^3(s){\rm exp}\lkk -\frac{1}{2}\bfk^2 R^2(s) \rkk
-R'^3(s){\rm exp}\lkk -\frac{1}{2}\bfk^2 R'^2(s) \rkk .
\end{equation}
%
In order to eliminate the radius dependence of the mass $M(R)$
we introduce new non-dimensional variables:
%
\begin{equation}
z:=\lmk \frac{R}{R_0} \rmk^{3/2}~~~~~~~~~{\rm and}~~~~~~~~\tau:=\frac{9}{4}
\frac{t}{R_0},
\end{equation}
%
where $R_0:=(R/M)^{1/2}$. Defining $U(z):=(4/9)R_0V(R)$, we obtain the
action,
%
\begin{equation}
S[z]:=\int d \tau \lkk \frac{1}{2}\lmk \frac{dz}{d \tau} \rmk^2
-U(z)\rkk.
\end{equation}
%
In the high temperature limit, the master equation becomes
%
\begin{eqnarray}
\frac{\ppp}{\ppp \tau} \rdm (z,z'; \tau) &  = &
\frac{i}{2}\frac{\ppp^2 \rdm}{\ppp z^2}
-\frac{i}{2}\frac{\ppp^2 \rdm}{\ppp z'^2}-iU(z)\rdm +iU(z') \rdm
  - \lmk\frac{2}{3} \rmk^2 \eta R_0 z_{\Delta}(\frac{\ppp}{\ppp
z}-\frac{\ppp}{\ppp z'})
\rdm
\nonumber \\
& - & \lmk \frac{2}{3} \rmk^4\frac{\Gamma bR_0^4}{16 \pi^{3/2}}
\lkk z^2+z'^2-2z^2z'^2 \lmk \frac{2}{z^{4/3}+z'^{4/3}} \rmk^{3/2} \rkk \rdm,
\end{eqnarray}
%
where $\eta \simeq (f^2T^2/48{\sqrt {2}}\Gamma)$ and
$b \simeq (f^2T^3 \phi_+^2/12 \Gamma^2)$. The derivation of
this master equation is tedious, but simple,
and is performed by the same procedure of Ref.\cite{cd}.
The last term in the right hand side is complicated compared
with the ordinary cases. However, only the region $z \simeq z'$ is
relevant for our purpose:
%
\begin{equation}
\lkk z^2+z'^3-2z^2z'^2 \lmk \frac{2}{z^{4/3}+z'^{4/3}} \rmk^{3/2} \rkk
\simeq \frac{1}{3}(z-z')^2+ {\cal O}\lmk (z-z')^4 \rmk.
\end{equation}
%
Thus we obtain the familiar result with small extra terms. These
extra terms might have appeared because
the ansatz of a subcritical bubble is not an exact solution
of the field equation.

\section{Friction, Diffusion and Decoherence}

In this section,we examine the validity of
the classical treatment for the evaluation of the typical scale of the thermal
fluctuation.
First we must give
the quantum decoherence condition. Here we define the following
quantity\cite{hiro};
%
\begin{equation}
\deco := {\rm Tr}(\rdm^2)
\end{equation}
%
This is a definite measure of classicality: $\deco$ becomes 1 for
the pure state and 0 if the quantum coherence is completely
destroyed.  This measure satisfies the equation
%
\begin{equation}
\frac{\ppp \deco}{\ppp t}=2 \eta  \deco -
\frac{4}{9}\frac{\Gamma b R_0^3}{8 \pi^{3/2}}
\int dz \int dz' \lkk z^2+z'^2-2z^2z'^2 \lmk \frac{2}{z^{4/3}+z'^{4/3}}
\rmk^{3/2} \rkk \rdm (z,z';\tau) \rdm(z',z;\tau).
\end{equation}
%
 From this one can read that the friction protects the
quantum coherence and the diffusion destroys it. The time
scales for these two effects are given by
%
\begin{equation}
\tau^{-1}_{\rm QC} \simeq 2\eta
\end{equation}
%
and
%
\begin{equation}
\tau^{-1}_{\rm QD} \simeq
\frac{4}{9}\frac{\Gamma b R_0^3}{8 \pi^{3/2}}\frac{1}{3}z^2 =
 \frac{2{\sqrt {2}}\eta T \phi_+^2}{27 \pi^{3/2}}R^3,
\end{equation}
%
respectively. $\qc$ is the time scale that the friction recovers
the quantum coherence and $\qd$ is that the diffusion destroys
the quantum coherence\footnote{The both time scales do not depend on
the coupling constant. This comes from the high temperature limit and
the fact that we take account into only the coupling with the one fermion.
If one consider the interaction with gauge fields, the dependence of
coupling constant appear. }.

For decoherence and complete classicalization, $\qd \leq \qc$ must
be satisfied and this inequality implies the lower bound for the
radius of a bubble,
%
\begin{equation}
R \geq \lmk \frac{27{\pi}^{3/2}}{{\sqrt {2}}T\phi_+^2} \rmk^{1/3} =:
R_1 \sim 0.084{\rm GeV}^{-1}.
\end{equation}
%
Furthermore one note that the life time of
subcritical bubbles should be longer than the time scale of
the complete quantum decoherence. The inequality
$\qd \leq \lt$ must be also satisfied, that is
%
\begin{equation}
R \geq \lmk \frac{27 \pi^{3/2}}{  {\sqrt {30}}\eta T \phi_+^2}
\rmk^{1/4} =: R_2 \sim 0.046{\rm GeV}^{-1}.
\end{equation}
%
Unfortunately, in the above argument it has been a simple
order estimations and therefore we cannot determine the exact
value for the critical size.
On the other hand the average radius is
$\langle R \rangle_T \sim 0.012{\rm GeV}^{-1}$.
Thus the critical size where quantum-to-classical transition occurs
is roughly given by $\langle R \rangle_T $.

The above result suggests that subcritical
bubbles should be treated by quantum mechanics and the
typical size should be calculated using the Wigner function.

\section{Summary and Discussion}

We estimated the critical size where quantum-to-classical transition
occurs. It turned out to be the same order as the
classical statistical averaged radius. This means that
subcritical bubbles should be treated as quantum systems with
dissipation at the critical temperature.

Although we have treated the fluctuations as classical in our
previous papers, at least in the minimal standard model with $m_H
=60$GeV, they are quantum rather than classical. Fortunately,
one can guess that
the quantitative result calculated based on quantum mechanics does not
have drastic change on the typical size(beside a factor of order one)
because the bubble is in the boundary between classical and quantum
region and then both results should be the same order with each
other. Hence we might conclude again that electroweak phase transition in
minimal standard model cannot accompany any supercooling
even if the potential is the first order type.

In order to obtain the exact value for the typical size of the
thermal fluctuation one must solve the master
equation or follow the time evolution of the Wigner function.
As one cannot take the limit $\hbar \to 0$ in the present problem,
the equation for the Wigner function does not coincide with
the classical Fokker-Planck equation obtained by reading the
imaginary part of the effective potential as noise in the previous
paper\cite{smy2}. Some higher derivative terms appear and left
for the case in which one cannot $\hbar \to 0$. Moreover,
despite the term (3.18) gives the diffusion term
in quantum Fokker-Planck equation, the corresponding noise
is not Gaussian as in the previous paper.
These problems will be investigated in our future study.

\vskip 1cm

\centerline{\bf Acknowledgment}
The author would like to thank H.\ Sato, M. Sasaki and
K. Nakao for their continuous encouragements and educational advises.
He also thanks M. Morikawa, J. Yokoyama, T. Tanaka
and S. Mukohyama for their discussions and careful readings of this
manuscript.
This work was supported by Grant-in-Aid
for Scientific Research Fellowship, No.\ 2925.


\end{document}